\documentclass[12pt,preprint]{aastex}
\usepackage{rotating}
\usepackage{changepage}
\usepackage{footnote}
\usepackage{footmisc}
\usepackage{amsmath}

\newcommand{\ltsim}{\raisebox{-1.0ex}{$\stackrel{\textstyle<}{\sim}$}}

\def\kms{km~s$^{-1}$}

\def\goes{{\sl GOES}}
\def\yohkoh{{\sl Yohkoh}}

\def\hinode{{\sl Hinode}}
\def\stereo{{\sl STEREO}}
\def\sdo{{\sl SDO}}

\def\p78{{\sl P78-1}}

\def\soho{{\sl SOHO}}

\def\halpha{H$\alpha$}

\def\kms{km~s$^{-1}$}

\def\etal{et~al.}

\begin{document}
%
\slugcomment{To appear in \underline{The Astrophysical Journal} (Accepted: 18 June 2018)}

\title{Magnetic Flux Cancelation as the Buildup and Trigger Mechanism for CME-Producing 
Eruptions in two Small Active Regions}

\author{Alphonse C.~Sterling\altaffilmark{1}, Ronald L. Moore\altaffilmark{1,2}, Navdeep K. Panesar\altaffilmark{1}}

\altaffiltext{1}{NASA Marshall Space Flight Center, Huntsville, AL 35812, USA,\newline alphonse.sterling@nasa.gov, ron.moore@nasa.gov}
\altaffiltext{2}{Center for Space Plasma and Aeronomic Research, University of Alabama in Huntsville, AL 35899, USA}


\begin{abstract}

We follow two small, magnetically isolated CME-producing solar active regions (ARs) from the
time of their  emergence until several days later, when their core regions erupt to produce 
the CMEs.  In both cases,  magnetograms show: (a) following an initial period where
the poles of the emerging regions separate from each other, the poles then reverse direction and
start to retract inward; (b) during the retraction period, flux cancelation  occurs along the
main neutral  line of the regions; (c) this cancelation  builds the sheared core field/flux rope
that eventually erupts to make the CME\@.  In  the two cases, respectively 30\% and 50\% of the
maximum flux of the region cancels prior to the eruption. Recent studies indicate that solar
coronal jets frequently result from small-scale  filaments eruptions, with those
``minifilament'' eruptions also being built up and triggered by cancelation of magnetic flux.
Together, the small-AR eruptions here and the coronal jet results suggest that isolated bipolar
regions tend to erupt when some threshold fraction, perhaps in the range of 50\%, of the 
region's maximum flux
has canceled. Our observed erupting filaments/flux ropes form at sites of flux cancelation, in
agreement with previous observations.  Thus,  the recent finding that minifilaments that erupt
to form jets also form via flux cancelation is further evidence that minifilaments are
small-scale versions of the long-studied  full-sized filaments.

\end{abstract}


\section{Introduction}
\label{sec-introduction}

Large-scale solar eruptions that produce solar flares, coronal mass ejections (CMEs), and
other  phenomena often start with the eruption of a filament structure of size  $3\times
10^4$---$1.1\times 10^5$~km \citet{bernasconi_et05}.  What builds these
filaments, and what eventually triggers them to undergo violent eruption, are long-standing
questions.  Because filaments form in magnetic field that runs along a magnetic neutral 
line, it is clear that magnetic free
energy, i.e.\ energy in the field in excess of that of a potential configuration, is a
requirement. Magnetic flux cancelation at the neutral line has been observed to precede
filament formation \citep{martin86}, and \citet{van_ball_et89} developed a theoretical
picture explaining how such  cancelation can result in a filament.  

Among the questions that remain is: what are the specific circumstances that determine when
the  filament will erupt to produce a flare and, if ejective, a CME\@?  It is a challenge to
address  this question with large-scale eruptions for the following reason: CME-producing 
filament eruptions often occur in active regions (ARs) that emerge and grow to a substantial
size before eruption takes  place; apparently, frequently the field must first evolve over
some time after the region  emerges, and so it can take two weeks or more for the AR to mature
and generate  a strong solar eruption. Therefore, because a complete solar rotation is
$\sim$27 days, it is often  not possible to observe the same flare-producing AR on the Sun
from the time that AR is born until the time of eruption.  That is, the AR in which a
large-scale eruption visible from Earth occurs usually was born when that region was on the far 
side of the Sun, if not even earlier.  This means that it is frequently not possible to follow in its
entirety the magnetic buildup of a  large-scale solar eruption with a set of near-Earth
instruments. This makes it difficult to observe the entire history leading up to the eruption,
and so our understanding of the eruption's buildup is frequently limited in
scope.  And, because the time-scale for buildup to eruption is comparatively long, it can be
difficult to separate out the key aspects that are most important for  understanding eruption
onset.  That is, if a strong-flare-producing AR develops on the far side of the Sun, we
could miss key  magnetic changes leading to an eruption in the AR when it is on the near
side.  Sorting out which  magnetic changes are key to an eruption are made more difficult in
such cases.  

It therefore would be advantageous if there were short-lived small magnetic active  regions that we
could study in their entirety during their disk passage, and over a short-enough period so that the
chance of complicating other substantial magnetic flux changes occurring during the build-up to a
whole-region-scale eruption would be minimized.  One such category of such  small active solar magnetic
regions are those whose whole-region-scale eruptions  make {\it solar coronal jets}. 
\citet{sterling_et15} present evidence that these jets are small-scale versions of larger-scale
eruptions \citep[for a similar view, see, e.g.,][] {hong_et11,hong_et14,shen_et11,innes_et13},
implying that lessons from jet-producing regions can inform us about larger-scale-eruption buildup and
triggering.

Seen in X-rays with \hinode/XRT, coronal jets are long (reaching $\sim$50,000~km), narrow 
($\sim$8,000~km), transient (lifetime $\sim$10~min) features that shoot  into the corona at
a  rate of $\sim$60/day in polar coronal holes \citep{cirtain_et07,savcheva_et07}, and a 
``jet bright  point'' (JBP) forms at an edge of the base of the jet.  Coronal jets are 
also visible in EUV coronal images \citep[e.g.][]{raouafi_et08,nistico09}, and they are
prevalent in quiet Sun and ARs in addition to coronal  holes
\citep[e.g.][]{shimojo_et96,raouafi_et16,panesar_et17,panesar_et18,sterling_et17,hong_et16,hong_et17}. 
Figure~1 shows a schematic picture for jet production presented by \citet{sterling_et15},
based on the idea that eruption of a small-scale filament, or ``minifilament,'' is the
source of the jets.  This figure is modified slightly from earlier versions
\citep{sterling_et15,sterling_et16,sterling_et17} with the addition of the red, dashed
reconnected field lines in
panel~(b), motivated by a recent study of \citet{moore_et18} demonstrating that
reconnection at the magnetic null point in the corona often begins earlier than or
concurrent with the tether-cutting reconnection below the rising minifilament.
This minifilament eruption model for jets has recently been successfully simulated
\citep{wyper_et17,wyper_et18}.

Based on recent observations, it is now clear that at least many coronal jets
are small-scale versions of large-scale eruptions 
\citep[e.g.][]{shen_et12a,adams_et14,sterling_et15, 
sterling_et16,sterling_et17,panesar_et16a, panesar_et16b,panesar_et17,panesar_et18}.  Seen
in profile at the solar limb, the minifilaments that erupt to cause the jets are of a much 
smaller size ($\sim$8000~km) but otherwise appear to be similar to  ``normal''-sized
filaments.

There are direct analogies between the coronal jets, and the  the large-scale magnetic
eruptions that make flares and CMEs: coronal jets result from eruptions of minifilaments,
generate  a miniature flare (the JBP), and eject material onto open field lines that guide the
jet (analogous to the CME in many large-scale eruptions) \citep{sterling_et15}. Besides
physical size, another difference between larger-scale eruptions and the eruptions that make
jets is that the buildup to eruption is much shorter with the jets: \citet{panesar_et17} found
that, for ten solar quiet region jets, the minifilaments that erupt to make jets
evolve from birth to eruption over periods ranging  from 1.5~hr to two days, and thus much
shorter than the two weeks or more for the ARs hosting large-scale eruptions to evolve and
release an explosive eruption.

Because the erupting minifilaments are analogous to the erupting larger-scale filaments,  we
suspect that the mechanism building and triggering the minifilaments to erupt is the same,
except for size scale (and magnitude), as the buildup and triggering mechanism  for
large-scale eruptions.  Minifilaments that erupt to form jets evolve rapidly enough so that we
can investigate directly the magnetic flux changes leading to their eruption.  Over several
investigations (\S\ref{sec-cancelation in jets}) we have now measured magnetic-change 
quantities of the fields at the base of a number of jets.  The object of the current study is
to examine the magnetic properties of a CME-producing region that is larger scale than
those that produce jets, for comparison with the situation in jets.  Our selected erupting
region however is small enough so that we can observe the region from the time of that field's
emergence through the time of eruption. This allows us to measure in this CME-producing
eruption the same magnetic quantities that we measured in the jets.

\section{Magnetic Cancelation in Jets}
\label{sec-cancelation in jets}

We have previously studied the properties of the flux cancelations around the time of jet
onset. \citeauthor{panesar_et16b}~(\citeyear{panesar_et16b}, \citeyear{panesar_et17}) 
examined 10-15 quiet-region jets (the actual number studied depends on how one counts 
repeated, a.k.a.\ homologous, jets); and \citet{panesar_et18} studied 13 coronal  hole jets. 
All of the quiet region and coronal hole jets either showed clear flux  cancelation at about
the time of the jet onset, or (in a minority of cases) the situation was  ambiguous but still
consistent with cancelation.  In the AR-jet cases \citep{sterling_et16,sterling_et17}, in only
one (out of $\approx$17 in total) jet episode \citep[in][]{sterling_et16} was it unclear
whether cancelation was occurring at the jet location near the time of jet onset; in all other
cases, cancelation was clearly present at the base of the jets at the time the jets were
occurring.

For the nine quiet region jets where reliable measurements were possible, 
\citet{panesar_et16b} found that, between a time substantially before the jet (5---6 hrs) and 
soon after the jet (0---1
hrs), the percentage of minority-flux reduction ranged between 18\%---57\%.  From the 
reduction percentage for each jet (which we call $p_i$) and their corresponding 
standard deviations ($\sigma_i$) listed in Table~1 of \citet{panesar_et16b}, 
we can calculate  a
weighted mean, $\overline{P}$, and weighted standard deviation, $\sigma_{\overline{P}}$, for 
the percent reduction in flux for each of those quiet region jets.  Using $1/\sigma_i^2$ as
weighting factors, we use for these quantities \citep[e.g.,][]{bevington_et03}: 

\begin{equation}
\overline{P} = \cfrac{\sum\limits_{i=1}^N p_i/\sigma_i^2}{\sum\limits_{n=1}^N 1/\sigma_i^2},
\end{equation}

and 

\begin{equation}
\sigma_{\overline{P}}^2 = \frac{N\sum\limits_{n=1}^N (p_i-\overline{P})^2/\sigma_i^2}{(N-1)\sum\limits_{n=1}^N 1/\sigma_i^2},
\end{equation}

\noindent
where $N$ is the number of jets examined.  This gives a weighted mean and weighted standard deviation
of 36.6\%$\pm 12.8$\% for the nine quiet Sun jets.  Also, for these quiet Sun jets \citet{panesar_et16b} 
found the minority flux patches generally to be of order $\sim$10$^{19}$~Mx.  

In a similar fashion, for coronal hole jets
\citet{panesar_et18} found flux reductions for 11 jets to range between 21\%---73\%.  Applying (1) 
and (2) to the values in Table~1 of \citet{panesar_et18} gives
weighted average and weighted standard deviation of 44.8\%$\pm 15.8$\%.  For these coronal 
hole jets,  \citet{panesar_et18} found the minority flux patches again 
generally to be of order $\sim$10$^{19}$~Mx, and they also calculated an average cancelation 
rate of $\sim$0.6$\times 10^{18}$~Mx/hr.

For approximately seven AR jets, \citet{sterling_et17} found on average  $\sim$5$\times
10^{18}$~Mx canceled prior to each jet episode, that the flux cancelation occurred at a rate
of $\sim$1.5$\times 10^{19}$~Mx/hr, and estimated that $\sim$10$^{28}$---$\sim$10$^{29}$ ergs
of free magnetic energy built up per jet. Due to the dynamic nature and polarity-arrangement 
complexity of the region however, it was not as straightforward as in the quiet Sun and 
coronal hole cases to measure the
percentage of flux  change in the buildups to the jets in the \citet{sterling_et17} study.  We
summarize various coronal jet properties in Table~1.

Several other detailed investigations of one or a small number of jets also
found them to originate from locations where magnetic flux cancelation occurs 
\citep{hong_et11,shen_et12a,huang_et12, adams_et14,young_et14a,young_et14b,shen_et17}.  Some other
jets however apparently occur in the absence of  obvious flux cancelation at the
HMI-detectable level \citep{kumar_et18}, although a study of 20  jets by
\citet{mulay_et16} also supports that most jets do occur at sites that include 
obvious flux cancelation.

Here we will study two CME-producing ARs for which we can track the entire magnetic 
history of the regions, from time of emergence until the time of the CME-producing
eruptions.

\section{Instrumentation and Data}
\label{sec-instruments}

For this study we use data from two instruments on the Solar Dynamics Observatory (SDO\@).  
That satellite's  Atmospheric Imaging Assembly \citep[AIA;][]{lemen_et12} images the Sun in
seven EUV wavelength bands: 304~\AA, 171~\AA, 193~\AA, 211~\AA, 131~\AA, 94~\AA, 335~\AA, in two
UV bands: 1600~\AA\ and 1700~\AA, and in a visible band: 4500~\AA\@. These channels respond to
plasmas emitting over a wide range of temperatures in the solar atmosphere \citep{lemen_et12},
observing  the full disk with $0''.6$ pixels and typically with cadence of 12~s and 24~s  in the
EUV and UV channels respectively.  Our magnetic data are from full-disk line-of-sight
magnetograms from \sdo's Helioseismic and Magnetic Imager \citep[HMI;][]{scherrer_et12}, with
45~s cadence and  $0''.5$ pixels.

For this study we searched for ARs that (a) erupt to produce a CME, (b) that we can follow
on the solar disk from emergence until the time of eruption within a single disk passage,  and
(c) that remain well isolated from other magnetic flux emergence until the time of eruption.  
Additionally, because we want to track the region as it builds up the  free energy in the
field that eventually erupts,  we avoided regions that  emerge already containing substantial
free energy concentrated in their cores and can explosively erupt shortly after  emergence,
such as is the case with some delta-sunspot ARs.

For our two selected regions, because of the short development time necessitated by condition
(b), in both cases the emerged AR is small and the resulting CME-producing  eruptions are
correspondingly small.  Our first example makes a CME-producing  eruption with a \goes\ C-class 
flare of 2013 October~20 from AR~11868.  This region had small sunspots visible on October~18.  Also, it
had two filament/flux ropes, one along the region's main neutral line that was faint in EUV, and
a second one along a neutral line on the outside edge of the region; as we will discuss below,
both filaments erupted in this region's eruption episode.  The second active  region that we
study produced a B-class  flare on 2010 July~16 from AR~11088; this region had a small sunspot
from early in its life that decayed over the course of our observations. This second region also had a
filament prominent in EUV running along its main neutral line, and this filament erupted during
the eruption episode.

In both cases the
resulting CMEs had widths smaller than many larger eruptions (as discussed below).  On the other hand,
in both cases
the erupting filaments had 2-D projected spans (i.e., not considering inclination angle on the
disk or curvature of the filament) of $\sim$70$''$; this is substantially longer than the
$\sim$8000~km ($\sim$11$''$) for  minifilaments measured in polar coronal hole jets by
\citet{sterling_et15}, or the jet-base widths for on-disk jets in coronal holes
($\sim$12{,}000~km)($\sim$17$''$) of \citet{panesar_et18} and in quiet regions
($\sim$17{,}000~km)($\sim$23$''$) of \citet{panesar_et16b}.  Therefore, we are observing
eruptions of a larger  category than those producing typical coronal jets.

\section{Eruption of 2013 October 20}
\label{sec-event1}

\subsection{EUV Development}
\label{subsec-event1_aia}

Figure~2 shows the eruption of 2013 October~20 in AIA~131 and AIA~304~\AA\ images at various
times, with Figures~2(d) and~2(h) showing overlaid HMI magnetograms.  In describing this 
eruption, we will also point out similarities with coronal jets.

Figures~2(a)---2(c) show that one of the two flux rope/filaments of this region erupts out from
the main neutral line (Fig.~2(d)), where that neutral line is apparent in the magnetograms of
Figures~2(d) and~2(h).   This flux rope/filament is not very prominent in EUV; while visible in
all seven AIA EUV channels, it is perhaps most obvious in the close-up 304~\AA\ images of
Figures~2(a)---2(c).  (At times leading up to eruption, this filament is prominent in GONG
\halpha\ images, available at gong.nso.edu.  Discussion of these images however is outside the
focus of the current paper.) The main flare brightening of the eruption occurs on the same main,
strong magnetic neutral line (Figs.~2(d) and~2(e)).  Comparing with the jet picture, this main
neutral line  corresponds to the neutral line defined by polarities M2 and M3 in Figure~1(a),
where the JBP develops in Figures~1(b) and~1(c)).  

Away  from that core location where the main brightening occurs, there is a bright rim that
partially surrounds the core erupting region; arrows in Figure~2(e) point out these remote
brightenings, which are also seen in  some ejective flares
\citep[e.g.][]{masson_et09,sun_et12,joshi_et15}; \citet{joshi_et17} have argued that
``three-ribbon flare'' events such as this are closely analogous to jets. Circular  ribbons such
as these, sometimes seen as only a portion of circles, are common in coronal jets (e.g.,
\citeauthor{li_et17}~\citeyear{li_et17}, \citeyear{li_et18};
\citeauthor{panesar_et16b}~\citeyear{panesar_et16b}, Fig.~1b), and also in chromospheric jets
known as surges (see discussion in \citeauthor{sterling_et16}~\citeyear{sterling_et16}, \S4). 
Comparing these circular brightenings with magnetograms (Fig.~2(h); also see Fig.~5(a)), it is
obvious that they are lodged against the edge of dominant-polarity (negative in this case)
field.  This agrees with the jet picture in Figure~1, where the circular ribbon is mapped out by
the negative-polarity footpoint of the  dashed-line reconnection loop in Figure~1(b). For the
eruption in Figure~2,  the loops in the large-scale event  corresponding to the Figure~1(b)
dashed loop itself are visible in the AIA 131~\AA\ video  corresponding to Figure~2 (many such
loops are obvious at, e.g., 09:23:44~UT in that video).

Our particular event in Figure~2 however is more  complicated than the basic jet picture of
Figure~1. In the core region, where the flare starts to brighten  from about 08:32~UT in the
video accompanying Figure~2, as noted above, in EUV images a well-developed filament does not
form and the flux rope that erupts is faint (Figs.~2(a)---2(c)).  There is, on the other hand, a
large filament that forms at and erupts from a location in-between the core (where the bright
flare is occurring in Fig.~2(b))  and the remote brightenings marked by the arrows in
Figure~2(e).  This filament begins to erupt at about 08:44~UT on 2013 October~20, which is after the
core eruption starts at about 08:34~UT\@. We discuss this filament eruption further in
\S\ref{subsec-filament}.

Figure~3 and the accompanying video show that this region's eruption produced a CME, 
visible in \soho/LASCO/C2 running-difference images (obtained from the 
online \soho/LASCO CME catalog; \citeauthor{gopal_et09}~\citeyear{gopal_et09}), 
first visible off of the solar west limb at 09:12~UT, and eventually having 
width $\sim$60$^\circ$. From the movie we estimate its speed to be $\sim$400~\kms. 
We suspect that the CME has 
components from the eruption of the core-region flux rope, as well as from the eruption 
of the prominent filament, with both eruptions visible in the videos accompanying Figure~2.  
It is not however possible to say with certainty whether one or the other of the eruptions, 
or both together, contributed most to the visible CME\@.

\subsection{Magnetic Field Evolution}
\label{subsec-b evolution}

From HMI magnetograms, AR~11868, from which  the 2013 October~20 eruption occurred, 
started emerging at about 14:00~UT on 2013 October~15. It underwent a surge in emergence from
about 04:00~UT on October~16, and an even  stronger surge from about 02:30~UT on October~18.  When it
started emerging on October~15 it was about $400''$ east of central meridian, and at the time of
eruption near 08:40~UT on October~20 it is $500''$ west of central meridian.  Therefore it is
an ideal example of a small bipolar AR with a relatively short evolution period of
about five days, before producing an eruption.  Moreover, the eruption was substantially
larger (in both geometric  size and EUV flux) than a coronal jet, and also drove a CME
of modest width.

Figure~4 shows the evolution of the magnetic region, where we have rotated all of the
images to a common time of 09:00~UT on 2013 October~20, and so close to the eruption time; all of
the other presented solar  images and movies for this event are rotated to this same
time.   In the images of Figure~4 we only show the region from after the start of the
strongest surge in emergence, while the movie accompanying that figure shows the region's
evolution over the entire time period, beginning before its emergence on October~14.

From Figure~4 and the accompanying movie, we see that the two polarities of the region
first separate from each other, and that separating motion continues until around 
03:00~UT on
October~19.  But then {\it the two polarities start to converge upon themselves.}  As that
convergence continues, the two poles of the bipole region start to interact and cancel, 
from about 03:00~UT on October~20, and this convergence and cancelation continues through the 
time of the eruption near 08:40~UT on October~20.

Figure~5 is a different representation of this dynamical evolution over the time
span covered by Figure~4, that is, from the start of the final burst of flux
emergence thorough to the time of the eruption. This is a time-distance map of that
evolution, with the ordinate values representing horizontal integrations across the
region covered by the white box in Figure~4(a), plotted as a function of time; the
vertical axis in Figure~5 therefore  represents distance along the y-direction in
the panels of Figure~4.  This clearly shows the  two flux polarities separating
early during the period, and then converging at later times,  with the eruption
occurring near the time when strong patches of the opposite polarities  collide.

Figure~6 shows the time evolution of the positive flux of the region, integrated
over the  area of the white box in Figure~4(a).   Although the HMI magnetograms that
we are using only  provide the line-of-sight component, we have approximately
accounted for foreshortening in the Earth-directed projection of that component by
dividing by the cosine  of the heliocentric angle between the AR's  location and
disk center (as observed by HMI),  as a function of time.  (When using non-derotated
magnetograms, there is an additional  cosine factor required to correct for
foreshortening of the area used in the flux calculation,  meaning that one should
divide by cosine squared instead of cosine.  Correction to this  area factor however
was accounted for with the derotation procedure that we used, which was  the
solarsoft \citep{freeland_et98} routine ``drot\_map,'' and therefore here we  only
need to divide by cosine to the first power.)  We only measure the positive-polarity
flux because it can be well isolated within  the box over the entire duration of the
measurement shown in Figure~6.  In contrast, the negative flux has obvious flows
across the Figure~4(a)-box boundary, and  hence cannot be measured reliably over
that region.  Our plot in Figure~6 covers the entire period from the beginning of
the flux emergence, showing the weak initial emergence near 14:00~UT on October~15, and
then the two subsequent  surges in emergence.  This shows that when eruption occurs
(orange line), the cancelation has been  going on for some time. 

Using the best-fit green line in the Figure~6 plot, we estimate the rate of flux
cancelation to be $1.1\times 10^{19}$~Mx/hr over the full time period of the green
line (30.75 hr; see Fig.~6 caption). If however we instead consider that the true
fall-off commences with  the bump in flux near 18:00~UT on October~19, then the rate
increases to $\sim$1.5$\times 10^{19}$~Mx/hr.  Therefore we take the rate of  flux
decrease to be $\sim$(1.3$\pm 0.2) \times 10^{19}$~Mx/hr.  If we assume that this
canceled  flux builds a magnetic flux rope, and  assuming equal amounts of positive
and negative flux cancel, this implies  $\sim$3.9$\times 10^{20}$~Mx of flux builds
up in the flux rope over a 30-hr period prior to the eruption.  (The arguments are
essentially unchanged if, instead of a flux rope, the cancelation goes  into
accumulation of non-potential shear along the neutral line.) From the plot in
Figure~6 we see that $\sim$30\% of the peak flux value canceled prior to the time of
the eruption.

We can estimate the amount of free energy contained in, say, the flux rope, formed by  the
canceled field. We can estimate the magnetic field strength, $B$, of the flux rope,  assuming
that the total accumulated flux passes through the approximate length-wise cross-sectional
area, $A$, of the flux rope. We estimate the size of the neutral line over which the
cancelation occurs to be $\sim$50$''$, based on the length of the abutting  positive and
negative fluxes in Figure~4(d)---4(f). Also, from Figure~4(f) we estimate, from the separation
of the two abutting polarities, that the width of the flux rope might be roughly $5''$; this
is about the separation distance between the positive and negative flux patches in
Figures~4(e) and~4(f), the latter image being about three hours after the eruption occurred.
We also  assume the height of the cancelation region to be the same as the width, $5''$.  So
the total accumulated flux determined above would be $BA = 3.9\times 10^{20}$, giving $B
\sim$300~G\@. With this value, and with the above dimensions for the flux rope, we can
estimate the amount of energy built  up in the flux rope volume, $V$ ($=A \times$~(width of
5$''$)).  Considering the roughness  in estimates for the flux rope size and in the magnetic
field strength, we estimate the accumulated energy$= V B^2/(8\pi)$ to be $\approx 2 \times
10^{30}$~erg.  With a wider flux rope and/or larger magnetic field strength the estimate could
be substantially  higher, and so we can say the accumulated energy is
$\sim$10$^{30}$---10$^{31}$~erg.

\subsection{Filament Formation and Ejection}
\label{subsec-filament}

We next examine the development of the second filament (the one that was prominent in EUV
images) that erupted with this event; in particular this question is of interest because, as
discussed above, the filament did not form and erupt from the main neutral line of the AR\@.
Looking at a 304~\AA\ movie of the AR from near the time of its emergence until time of eruption
(video accompanying Fig.~2) shows that the filament in the region develops over about 10---16~UT
on October~18, which is about two days prior to eruption.  From  magnetograms overlaid 
onto those images (Figs.~2(e) and~2(h)) and the
accompanying video (and also the magnetogram video accompanying Fig.~4), it is apparent that the
filament forms where positive-polarity flux elements outside and to the west of the positive
polarity pole of the emerging flux,  canceled with surrounding negative-polarity field over
about that same time period.  From magnetogram images from about the time of Figure~4(b), we
measure the filament-forming canceling elements to be of moderate strength, of
$\sim$100---300~G; arrows in Figure~4(b) point to some of these flux elements.  These values are
substantially weaker than the strongest-field portions of the emerging region at that  time,
which have strengths ranging from $\sim$500~G and up to and exceeding 1000~G\@. Thus it is the
intermediate-strength clumps that cancel to make the filament, rather than the strongest-field 
elements.  These observations of flux cancelation as the mechanism for formation of the
filament  are fully consistent with previous observational and theoretical studies
\citep[e.g.][]{martin86,van_ball_et89}. The canceling fluxes that form the filament are not
concentrated enough to allow for easy isolation and flux-buildup and flux-rate measurements, as
we did above with the canceling fluxes in the core region, and therefore we do not attempt such
measurements here.

Eruption of the filament begins with its liftoff from the surface, which starts at about 
08:41~UT on October~20; this is clearly later than the 08:32~UT start of the flare in the core
region.   Moreover, as the filament begins lifting off, remote ribbons of the region (arrows
in Fig.~2(e))  are also already illuminated.  This means that new loops are being added to the
lobe overlaying that filament, as at that point the eruption is in a situation analogous to
Figure~1(b) in the jet picture (where the filament of this October~20 eruption initially sits
beneath the central, larger lobe rooted in polarities M1 and M2 of Fig.~1(a)).  We suspect 
that eruption of the filament is nonetheless triggered by the disruption resulting from the
flux rope eruption along the main neutral line.  Although it might seem counterintuitive to
have the filament erupt through an overlying lobe that contains  newly closed field (the
dashed red loop in Fig.~1(b)), we have seen evidence before \citep{sterling_et14} for where a
strong second-stage filament erupted, apparently  bursting through (or partially navigating
around) newly closed field resulting  from a first-stage eruption.  \citet{torok_et11} 
present another example of eruption through newly-closed field, along with numerical
simulations demonstrating the plausibility  of such complex eruptions. A likely possibility is
that implosion of the field around the erupting core-region flux rope triggers the eruption of
the remote filament, the so-called ``Hudson Effect,'' from the ideas presented in
\citet{hudson00} \citep{janse_et07,shen_et12b,panesar_et13}.

\section{Eruption of 2010 July 16}
\label{sec-event2}

Our second example is of AR~11088, which emerged on 2010 July~11 and  resulted in a \goes\
B-class flare five days later on July~16. Figure~7 shows AIA 171~\AA\ and 304~\AA\ images of this event. 
In this case a filament erupted  from the main neutral line of the region, and was directed 
toward Earth, and hence  toward AIA\@. From the accompanying video, the eruption begins 
near 15:00~UT,  with strong flare emission apparent about 20~min later.

Figure~8 and the accompanying video show that a CME that likely originated from this eruption
was visible in \stereo-B COR~1 images (from the online COR1 CME catalog:
cor1.gsfc.nasa.gov/catalog/),  beginning from 15:41:02~UT on 2010 July~16, and, from
Figure~8(b),  obtaining a width of approximately 35$^\circ$. From this movie we estimate 
its speed to be $\sim$250~\kms. \stereo-A COR~1 shows what is likely the same CME from 
16:30:18~UT\@.  \soho/LASCO does not show a clear CME from this region, likely due to the 
CME's small width and on-disk location not far from disk center from the LASCO perspective.

Figure~9 shows the magnetic evolution of AR~11088, the source of this eruption, over the
five-day period from the region's emergence until the eruption.  Generally, the behavior is
similar to that of the 2013 event of \S\ref{sec-event1}, with the region emerging and then the
poles of the region separating, and then the poles retracting in on themselves, with
cancelation occurring at the location of the main neutral line. By the time of the eruption
(Fig.~9(f)), the flux has clearly declined markedly from the earlier displayed times.

Figure~10 is analogous to Figure~5, showing in a time-distance map the evolution of the positive and
negative fluxes with time.  We created the map by integrating along the north-south direction in the box
of Figure~9, and so in this case the vertical axis of Figure~10 represents distance along the
$x$-direction in the panels of Figure~9.  This shows that, although a substantial part of the flux of
the region still  remains, a substantial portion of the emerged flux also has undergone cancelation by the
time of the eruption (orange line).

Figure~11 shows quantitatively the amount of positive-flux reduction over the life of the
region, where we have again selected only the positive flux because it is (somewhat) easier to
isolate than the negative flux of this region, where we have integrated that positive flux as a
function of time over the location of the white box of Figure~9(a). Again we have divided by the
cosine of the  region's angle with disk center as a function of time.  As with the previous
event, there again is an overall  trend of flux reduction with time.  Inspection of AIA
1600~\AA, 1700~\AA, and 4500~\AA\ images together with the magnetograms shows that a small
positive-polarity spot appears near the time of emergence on July~11, and while it fades to a
small pore by the end of July~12, HMI magnetograms show that the concentrated positive flux of
the spot continues to disperse over most of the period covered by Figure~11; cancelation of
this flux apparently contributes to the steady decrease 
in positive flux of the region in Figure~11.

Between  the start of the decline at about 20~UT on July~11 and the
eruption onset time of 15:00~UT on July~16,  the flux drops from  $8.2 \times 10^{20}$~Mx to
$4.0 \times 10^{20}$~Mx, so the total reduction is $4.2 \times 10^{20}$~Mx.  Thus, over the
115-hr time period this is a rate of $\sim$3.7$ \times10^{18}$~Mx/hr (and we assume an uncertainty 
of about the same level as in the first event: $0.2\times10^{18}$~Mx/hr), and the total decrease
is about 50\% of the peak value.

Again we can estimate the energy built up in the expected flux rope (or sheared field) resulting
from the cancelation, again assuming that the canceled flux  ($4.2 \times 10^{20}$~Mx) goes into
a flux rope.  Using magnetograms to estimate the flux 
rope size, from Figure~9(e) we take its
length to be $40''$, and we again use a width of $5''$.  Following the procedure in
\S\ref{subsec-b evolution}, here we obtain $B\sim$400~G\@.  These values result in an estimate
for the energy, $V B^2/(8\pi)$, to be  $2.4 \times 10^{30}$~erg, and so similar to that of the
2013 event. With the previous provisos, we again conclude 
that $\sim$10$^{30}$---10$^{31}$~erg
of energy could have been accumulated from the cancelation.

From the long-duration 304~\AA\ movie accompanying Figure~7, we see that the filament of
the region began  to form from about 10:00~UT on July~14 (again about two days prior to eruption), 
with increased rate of formation from
about 12:45~UT on July~15, and full formation by 00:00~UT on July~16.  Due to the data dropout over
the second-half of July~14, what happens with the magnetic field during that time period is not
clear in Figure~10; that figure does however show  cancelation at the bipole's main neutral
line occurring during the second-half of July~15, consistent with that cancelation leading to
the filament's formation.  Similarly, cancelation is apparent along that neutral line in the
long-duration 304~\AA\ movie with overlaid magnetograms accompanying Figure~7, from about
22:00~UT on July~15 following a data gap.

\section{Summary and Discussion}
\label{sec-discussion}

We have followed two different small ARs, from the time that they emerged near the solar east
limb, until a time when they were the source for CME-producing eruptions.  In both cases: (1)
the ARs emerged, (2) their poles moved away from each other until reaching a  maximum
separation, (3) after that the poles started to retract in on themselves (shown well in Figs.~5
and~10), (4) after some time this retraction led  to magnetic flux cancelation along the main
neutral line of the emerged bipole, and (5) eventually the continued cancelation led to an
eruption along the main neutral line  of the retracted bipole.  In the 2013 AR of
\S\ref{sec-event1}, some of the positive flux on the outskirts of the emerging flux region
canceled with nearby negative-polarity flux, and the prominent EUV filament of the region formed
at that location, which was east of the main (central) neutral line of the  AR\@.  In that case,
the first eruption was of a (faint in EUV) flux rope that formed along the main neutral line as
a result of the cancelation of the retracting region, and that blowout core eruption apparently
triggered eruption of the prominent EUV filament a few minutes later, plausibly via the Hudson
effect.  In the 2010 AR of \S\ref{sec-event2}, a filament developed along the  main (central)
neutral line of the emerged/retracted AR, and it was that filament that erupted to generate the
CME\@.

Figures~3(b) and~7(b) show that the CMEs from these two small AR eruptions had widths  of
$\sim$60$^\circ$ and 35$^\circ$. Not all coronal jets produce ejections observable in
coronagraphs, but when they do,  typically coronal jets produce coronal ejections that are
substantially narrower than this, being $\ltsim 10^\circ$, and called ``white-light jets'' or
``narrow CMEs'' 
\citep[e.g.,][]{wang_et98,wang_et02,nistico09,paraschiv_et10,hong_et11,yu_et14,moore_et15,sterling_et16}.
In some cases, jets can trigger wider CMEs to erupt, when the magnetic field that erupts to drive those
jets, i.e.\ the erupting minifilament (or, presumably an ``empty minifilament field'' (flux
rope) alone, if there is no cool minifilament material on that field), erupts into the base of a
coronal loop and drives that loop out (via its magnetic twist transferred to the loop) to  form
a CME wider than 10$^\circ$. Because that larger coronal loop often forms the base of a coronal
streamer, we sometimes call these ``streamer puff'' CMEs \citep{bemporad_et05,panesar_et16a}. 
(\citeauthor{shen_et12a}~\citeyear{shen_et12a} also observe a ``bubble-like'' CME originating
from a jet region, and that event might also form as suggested in 
\citeauthor{panesar_et16a}~\citeyear{panesar_et16a}.)
In the two small-AR CMEs studied here, the eruption leading to the CMEs originates along the
main neutral line of the ARs, rather than off to the side of the main neutral line of the
blown-out loops as is  characteristic of streamer puff CMEs.  Therefore, the CMEs produced by
the two small ARs of this study differ from the narrow CMEs or the streamer puff CMEs that occur
with jets.  At the same time, from a study by \citet{gopal_et09} of over 10{,}000 CMEs, the
average CME width is 44$^\circ$, but with a large number wider then 60$^\circ$ \citep[see
Fig.~10 of][]{gopal_et09}; hence, the small ARs of this study produced CMEs of relatively
moderate width.  This is consistent with the idea that the small AR eruptions of this study are
on a continuum of solar eruptions occurring on different size scales, with coronal jets being
smaller-sized eruptions, and many large-scale eruptions being bigger than those presented here. 
This also agrees with the trend reported by \citet{moore_et07} for a direct proportionality
between magnetic strength of  the area from which the eruption occurs and the final width of the
ejected CME\@.

An initial motivation for this investigation was our previous conclusions that many if not
all coronal jets are  miniature versions of larger-scale CME-producing eruptions.  If this 
is true, then studying the triggering mechanism of coronal jets (which develop on a short
time scale of a couple days or less) could inform us of the triggering mechanism of
larger-scale eruptions (which develop on longer timescales,  sometimes one or more solar
rotations).  We found strong evidence that magnetic flux cancelation builds and 
triggers the CME-producing eruptions for both of our studied small ARs, supporting the
suggestion that coronal jets are useful proxies for studying the onset of larger-scale
eruptions.

Additionally, we found evidence that filaments in both of our small ARs formed via flux
cancelation. This has been found for large-scale filaments previously (see
\S\ref{sec-introduction}).  More recent observations show that flux cancelation
also builds minifilaments that erupt to drive jets \citep{panesar_et17}.  Thus, our results
here regarding filaments forming at cancelation sites also supports that there is a continuum
of similarly-produced filament-like features on a variety of size scales that erupt.

We also note that AR filaments can form in a variety of ways, for example, shearing 
motion and sunspot rotation \citep{yan_et15} and complex magnetic changes 
\citep[e.g.][]{schmieder_et04}.  Moreover, high-resolution investigations of formation 
and evolution of filaments show a wealth of complexities \citep[e.g.][]{berger_et08,shen_et15,
yang_et16a,yang_et16b}.  Details such as these regarding filament formation are beyond
the scope of the current paper.

The two small eruptions of this study do show some differences from the minifilaments that
erupt to produce jets; among these is the source of the cancelation that produces the
eruptions of this study, compared to the minifilaments in the jet studies. 
\citet{panesar_et17} found that, for the cases where they could make a determination, the
eventually-erupting minifilaments form (a) when tiny grains of flux coalesce to form a
minority polarity in a background majority-polarity region; or (b) the minority-polarity pole
of a small emerging bipole forms a bipole with a surrounding majority-polarity flux clump, and
the minifilament forms along the neutral line of that bipole.  In both of the cases presented
here, the core erupting feature (a faint flux rope in the 2013 case, and a well-defined
filament/flux rope in the 2010 case) both formed when the two poles of the emerged AR
retracted in upon themselves, and underwent flux cancelation  at the main neutral line of that
same previously-emerged bipole.  For the AR jets studied by \citet{sterling_et17}, the jets 
occurred in a homologous fashion according to scenario (b).  For the AR jets of 
\citet{sterling_et16} however, at least one of the jets (see \S3.5 of
\citeauthor{sterling_et16}~\citeyear{sterling_et16}) appears to have occurred along the 
internal neutral line of a sheared-core emerged bipole, similar to the two events of the
present paper.

Because we suspect that the events here are larger-scale versions of coronal jets, and because
both the CME-producing eruptions of this paper and many jets result from flux cancelation, we
can compare our flux changes for the two events  here with those of coronal jets. We present
these comparisons in Table~1 for coronal hole (CH) jets, quiet Sun (QS) jets, AR jets, and the
two small CME-producing eruptions of this paper.  From this table, it appears  that the {\it
rate} of flux cancelation does not seem to be a determinate for the size of the eruption that
will occur, since the 2010 small-AR eruption made an eruption apparently more energetic
($10^{30}$---$10^{31}$ erg) than the AR jets of  \citet{sterling_et17} ($10^{28}$---$10^{29}$
erg), while the rate of flux cancelation in that 2010 small AR ($4\times 10^{18}$~Mx/hr) was
smaller than that of those AR jets ($1.5\times 10^{19}$~Mx/hr).

Rather, the results suggest instead that the total {\it percentage} of flux of the entire region
that cancels may be more indicative of when an eruption may occur.  Although the range of values 
is large for the jets and the number of AR studied is small, currently the data indicate that by the time
$\sim$50\% of the flux of an isolated bipolar region has canceled, that region will either have
already erupted or eruption is imminent.  This suggests a physical explanation for the possible
relationship between flux-cancelation amount and eruption onset: it could be that it is
necessary for about 50\% of the region's total flux to cancel for enough free energy to be built
up in the flux rope (or sheared field), together with the field restraining the eruption (the
non-canceled field) to be weak enough, for the eruption to take place. This suggests that, at
least for regions that are sufficiently magnetically isolated from their surroundings,
monitoring the change in the amount of total flux of a potentially erupting region --- up to and
including large CME-producing regions --- could provide a prognosticator for when eruption will
eventually take place.  It will however be necessary to investigate additional events before 
drawing firm conclusions regarding these speculations.

\acknowledgments

We thank an anonymous referee for helpful suggestions, and also for many reference additions.  A.C.S. 
and R.L.M. were supported by funding from the Heliophysics Division of NASA's Science
Mission Directorate through the Heliophysics Guest Investigators (HGI) Program, and the
\hinode\ project.  N.K.P's research was supported by an appointment to the NASA Postdoctoral
Program at NASA MSFC, administered by Universities Space Research Association under contract
with NASA\@.

\clearpage
\setlength{\footnotesep}{1mm} 
\renewcommand{\hangfootparskip}{0ex}
\renewcommand{\hangfootparindent}{1em}

\begin{table}[ht]
\begin{adjustwidth}{0cm}{}
\caption{Magnetic properties for coronal jets and for CME-producing small ARs:}
\renewcommand{\arraystretch}{0.}
\vspace{6mm}
{\tiny

\def\arraystretch{1.1}	
\begin{minipage}{\textwidth}
\begin{tabular}{cccccc}

\hline\hline
Objects & Study  &  Cancel Rate ($10^{18}$~Mx/hr) & Canceled amount$^{(a)}$ ($10^{18}$~Mx) & Percentage & Energy (erg) \\
\noalign{\smallskip}\hline \noalign{\smallskip}

CH Jets  &  \citet{pucci_et13} & ---$^{(b)}$ &  ---$^{(b)}$  & ---$^{(b)}$ & $10^{26}$---$10^{27}$  \\   
CH Jets &  \citet{panesar_et18} & 0.6 &  0.5---2.0  & 45$\pm 16^{(c)}$ & ---$^{(b)}$  \\   
QS Jets &  \citet{panesar_et16b} & 1.5$^{(d)}$ &  0.9---4.0  & 37$\pm 13^{(e)}$ & ---$^{(b)}$  \\   
QS \& AR Jets$^{(f)}$ &  \citet{shimojo_et00} & ---$^{(b)}$ &  ---$^{(b)}$  & ---$^{(b)}$ & $10^{27}$---$10^{29}$  \\   
AR Jets &  \citet*{sterling_et17} & 15 &  5  & --- & $10^{28}$---$10^{29}$  \\   
2013 Oct~20 &  This paper, \S\ref{sec-event1} & 13 &  390  & $29\pm 3^{(g)}$ & $\sim$10$^{30}$---$10^{31}$ \\   
2010 Jul~16 &  This paper, \S\ref{sec-event2} & 4 &  420  & $51\pm 3^{(g)}$ & $\sim$10$^{30}$---$10^{31}$ \\   

\hline
\end{tabular}

\end{minipage}
\label{tab:list}

\vspace{0.6cm}
\noindent
(a) Total average flux canceled from start of flux decline, or from time of previous events in the case of homologous jets. \\
\vspace{-4mm}

(b) This quantity difficult to measure reliably, or otherwise not provided in stated study. \\
\vspace{-4mm}

(c) Determined using values in Table~1 of \citet{panesar_et18}; see text \S\ref{sec-cancelation in jets}. \\
\vspace{-4mm}

(d) Data for this value are from \citet{panesar_et16b}, but the calculation is described in \S4.1 of \citet{panesar_et18}.\\
\vspace{-4mm}

(e) Determined using values in Table~1 of \citet{panesar_et16b}; see text \S\ref{sec-cancelation in jets}. \\
\vspace{-4mm}

(f) Although this study examined jets in all regions, it used \yohkoh/SXT data, which were biased toward AR and QS
jets over the\newline comparatively-softer-X-ray CH jets.

(g) Uncertainty determined assuming assuming flux values accurate to $2 \times 10^{19}$~Mx.

} 
 \end{adjustwidth}
\end{table}
\clearpage

\figcaption{Schematic showing jet generation via a ``minifilament eruption model,'' as proposed in
\citet{sterling_et15} (with an adjustment due to
\citeauthor*{moore_et18}~\citeyear{moore_et18}; see text).  (a) Cross-sectional view of a 3D
positive-polarity anemone-type field inside of a majority negative-polarity ambient background
field (which is either open or far-reaching field).  One side of the anemone is highly sheared
and contains a minifilament (blue circle).  (b) Here the minifilament is erupting and
undergoing reconnection in two locations:  {\it internal} (``tether-cutting'' type)
reconnection (larger red X), with the solid red lines showing the resulting reconnected
fields; the thick red semicircle represents the ``jet-base bright point'' (JBP\@). {\it
External} (a.k.a.\ ``interchange'' or ``breakout'' reconnection) occurs at the site of the
smaller red X, with the dashed  lines indicating its two reconnection products.  (c) If the
external reconnection proceeds far enough, then the minifilament material can leak out onto
the open/far-reaching field.  Shaded areas represent heated jet material visible in X-rays and
some \sdo/AIA EUV channels.  See captions of e.g.\ \citet{sterling_et15} or
\citet*{moore_et18} for a more detailed description. Labels M1, M2, and M3 are referred to in
the text; they respectively  point out negative, positive, and negative photospheric polarity
locations. \label{fig1}}

\figcaption{AIA~304~\AA\ (a--c; g--i), and 131~\AA\ (d--f) images of a small-AR  CME-producing
eruption, from 2013 October.   Contours in (d) are from an \sdo/HMI magnetogram from 08:30~UT on
October~20, and contours in (h) are from 16:45~UT on October~18, with red and yellow contours
respectively representing positive  and negative field.  Arrows in (a---c) point out a faint 
flux rope erupting away from the core of the region, where the main neutral line (d) is located.
Overall, the basic eruption geometry is analogous to that of the coronal jet in Fig.~1, where
the flare brightening along the neutral line in (d) corresponds to the JBP in Figs.~1(b)
and~1(c), and where the arrows in (e) point to  ``external brightenings,'' corresponding to the
footpoints of reconnected loop fields corresponding to the location marked M1 in Fig.~1(a).
Arrows in (h) point to a filament that forms along a neutral line  away from the main central
neutral line; this filament is erupting in (f) and (i).   Animations of  this figure are
available.  There are separate 304~\AA\ animations for (a--c) and for (d--f), with the former
being of shorter duration, higher cadence, closer-in field of view, and different color and
intensity scalings than the latter, with parameters of the former tuned to facilitate viewing of
the faint flux rope.  In these and all other solar images in this paper, north is up and west is
to the right.  For the event shown here, and in all other figures for this 2013 small-AR event,
the images were differentially rotated to the common time of 2013 October~20, 09:00~UT\@.
\label{fig2}}

\figcaption{\soho/LASCO/C2 running difference image of the CME accompanying the eruption of
Fig.~2. White lines in (b) subtend an angle of $\approx 60^\circ$.  An animation of this
figure is  available. \label{fig3}}

\figcaption{Magnetic flux changes from HMI of the region of Fig.~2, with white/black 
representing positive/negative fluxes,
respectively.  The box in (a) shows the region used to produce the
flux plot in Fig~5. Fluxes saturate at $\pm$300~G\@. Arrows in (b)  show examples of
mid-strength flux elements that cancel to form the filament that erupts in Fig.~2(f).
An animation of this figure is available. \label{fig4}}

\figcaption{Time-distance map for the event of Fig.~2 for the flux summed in the horizontal
(E-W) direction (i.e., in pixel rows) over the box of Fig.~4(a).  Flux emerges near the start; the two poles
spread apart, reaching a maximum separation near 0~UT on 19 Oct; and then the two poles 
converge back on themselves, with cancelation occurring at the central neutral line of the
bipole from approximately the start of October~20.  A CME-producing eruption occurred near 
08:00~UT on October~20.  Blue dashed lines indicate 00:00~UT for each day, and the solid orange line
shows the time of the eruption. \label{fig5}}

\figcaption{Variation with time of positive flux for the event of Fig.~2, where the flux is the
integrated  positive-polarity flux over the box in Fig.~4(a).  Values are calculated assuming a
field component vertical to the surface, where we have approximately corrected for
foreshortening by dividing by the cosine of the angle between the region and observed disk
center (see text).  The green line is a least-square fit over the period of the decay until the 
eruption (from 19 October 05:30~UT until 20 October 11:45~UT), and the orange line marks the
onset time of  the flare/CME-producing eruption. \label{fig6}}

\figcaption{AIA~\AA\ and 304~\AA\ images of a second small CME-producing eruption of this
study, this one from July 2010.  Contours in (a) are from an HMI magnetogram from 15:15~UT on
July~16, and contours in (e) are from 09:45~UT on the same day, with red and blue contours
respectively representing positive  and negative field.  In this case the filament is erupting
nearly directly out of the figure; arrows in (b) and in (e) point to the filament.  Animations
of  this figure are available.  Jumps in time during the last half of July~14 in the animations
are due to \sdo\ spacecraft pointing changes.  For the event shown here, and in all other
figures for this 2010 small-AR event, the images were differentially rotated to the common
time of 2010 July~16, 15:30~UT\@. \label{fig7}}

\figcaption{\stereo/COR1 running difference image of the CME accompanying the eruption of
Fig.~7. White lines in (b) subtend an angle of $\approx 35^\circ$.  An animation of this
figure is  available. \label{fig8}}

\figcaption{As in Fig.~4, but for the event of Fig.~7.  The box in (a) shows the region 
used to produce the flux plot in Fig.~9. Fluxes saturate at $\pm$300~G\@.  An animation 
of this figure is available. Jumps in time during the last half of July~14 in the animation 
are due to \sdo\ spacecraft pointing changes.  \label{fig9}}

\figcaption{Time-distance map, as in Fig.~5 but for the event of Fig.~7, where the flux
is summed along pixel columns in the vertical (N-S) direction of the box of Fig.~7(a).  The vertical strip
during the last half of July~14 is an artifact resulting from \sdo\ spacecraft operations
during that period, and so that portion of the map should be ignored.   A CME-producing
eruption occurred near 15:00~UT on July~16; the solid orange line shows the eruption time.
\label{fig10}}

\figcaption{Variation with time of positive flux for the event of Fig.~7, where the positive
flux  is integrated over the box in Fig.~7(a).  Values are calculated assuming a field
component vertical to the surface, where we have approximately corrected for foreshortening by
dividing by the cosine of the angle between the region and observed disk center (see text). 
The green line is a least-square fit over the period of the decay until the  eruption (from 
19 October 05:30~UT until 20 October 11:45~UT), and the orange line marks the onset time of the
flare/CME-producing eruption. \label{fig11}}

\clearpage
\pagestyle{empty}
\begin{figure}
\vspace{-5cm}
\hspace*{-4cm}\includegraphics[angle=-90,width=25cm,scale=1.1]{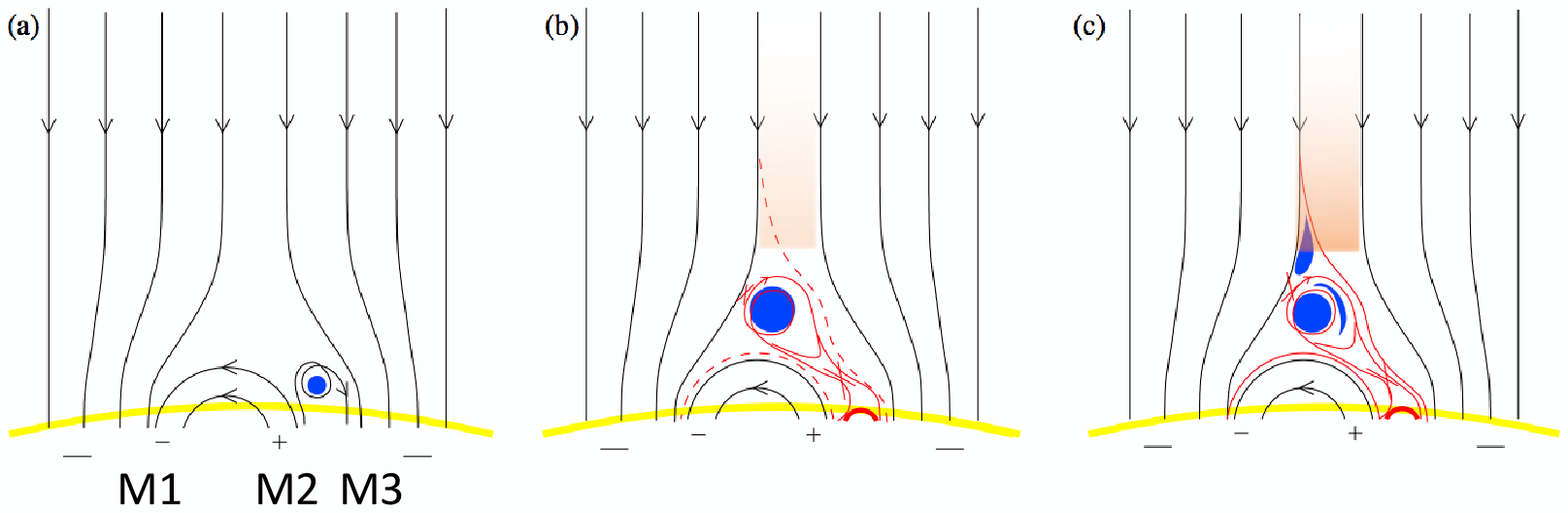}\vspace{-3cm} 
\centerline{Figure~1}
\end{figure}
\clearpage

\begin{figure}
\hspace*{0cm}\includegraphics[angle=0,scale=0.9]{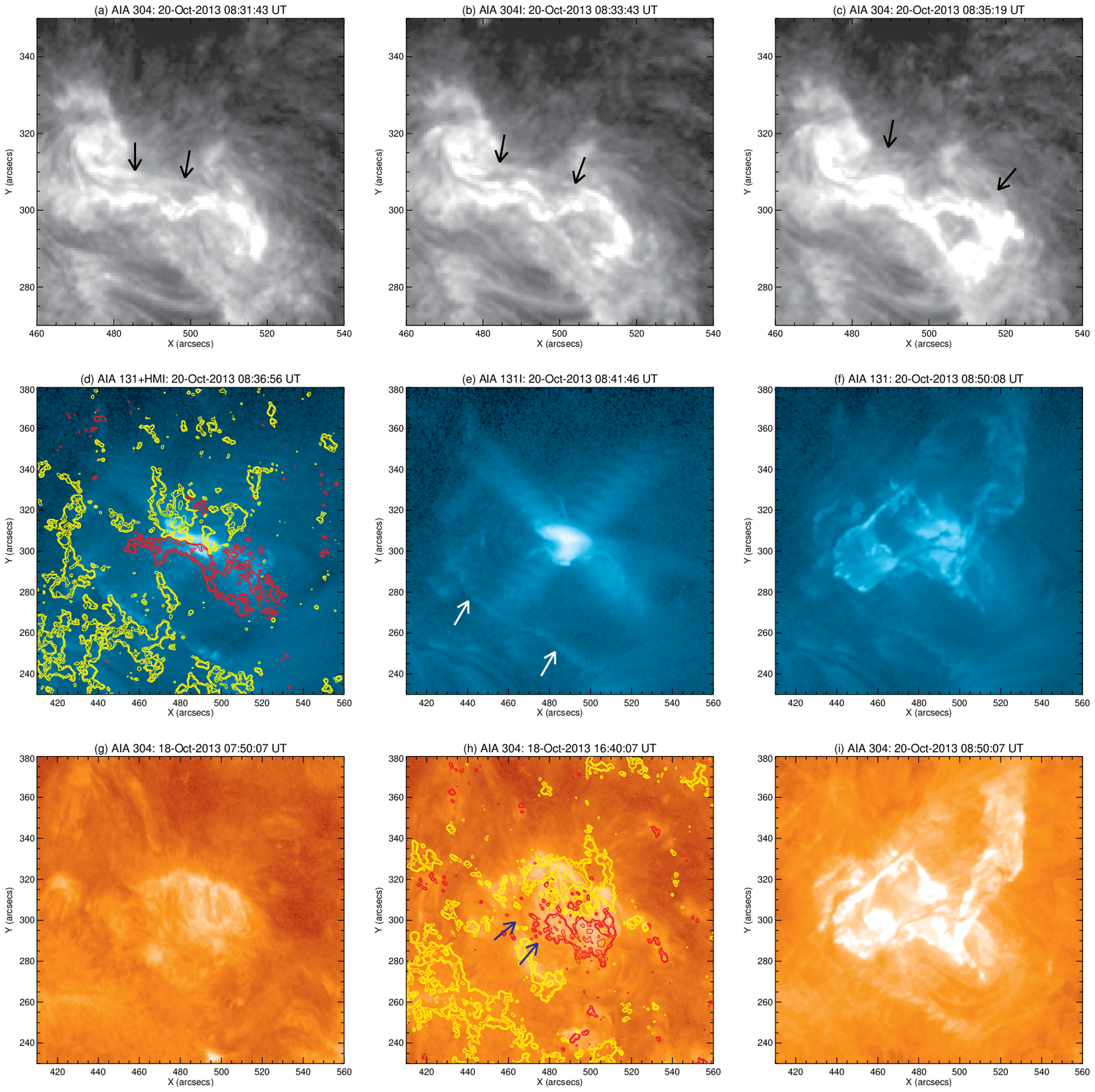}
\centerline{Figure~2}
\end{figure}
\clearpage

\clearpage
\pagestyle{empty}
\begin{figure}
\hspace*{0.5cm}\includegraphics[angle=0,scale=0.5]{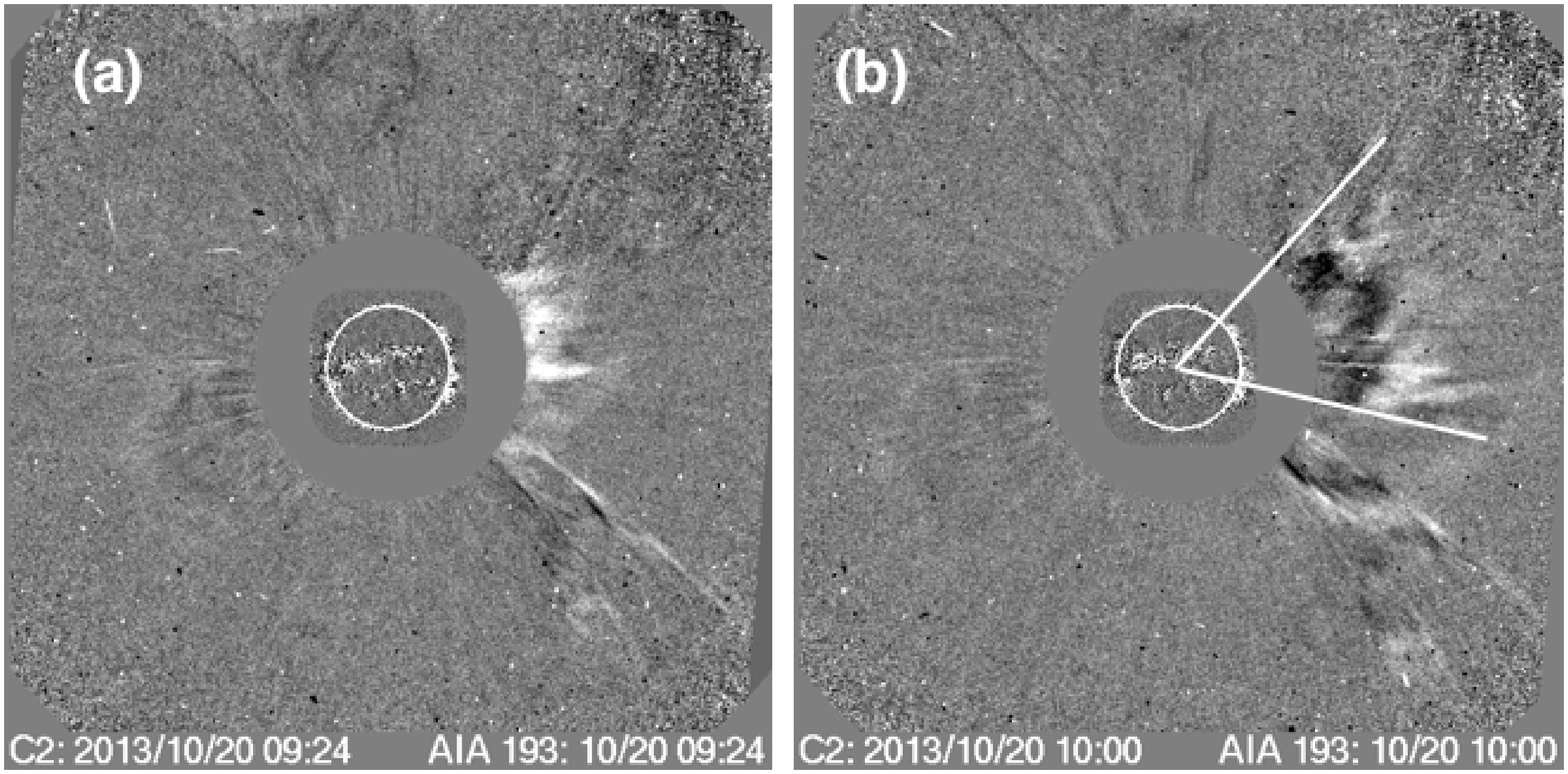}\vspace{1cm}	
\centerline{Figure~3}
\end{figure}
\clearpage

\clearpage
\pagestyle{empty}
\begin{figure}
\plotone{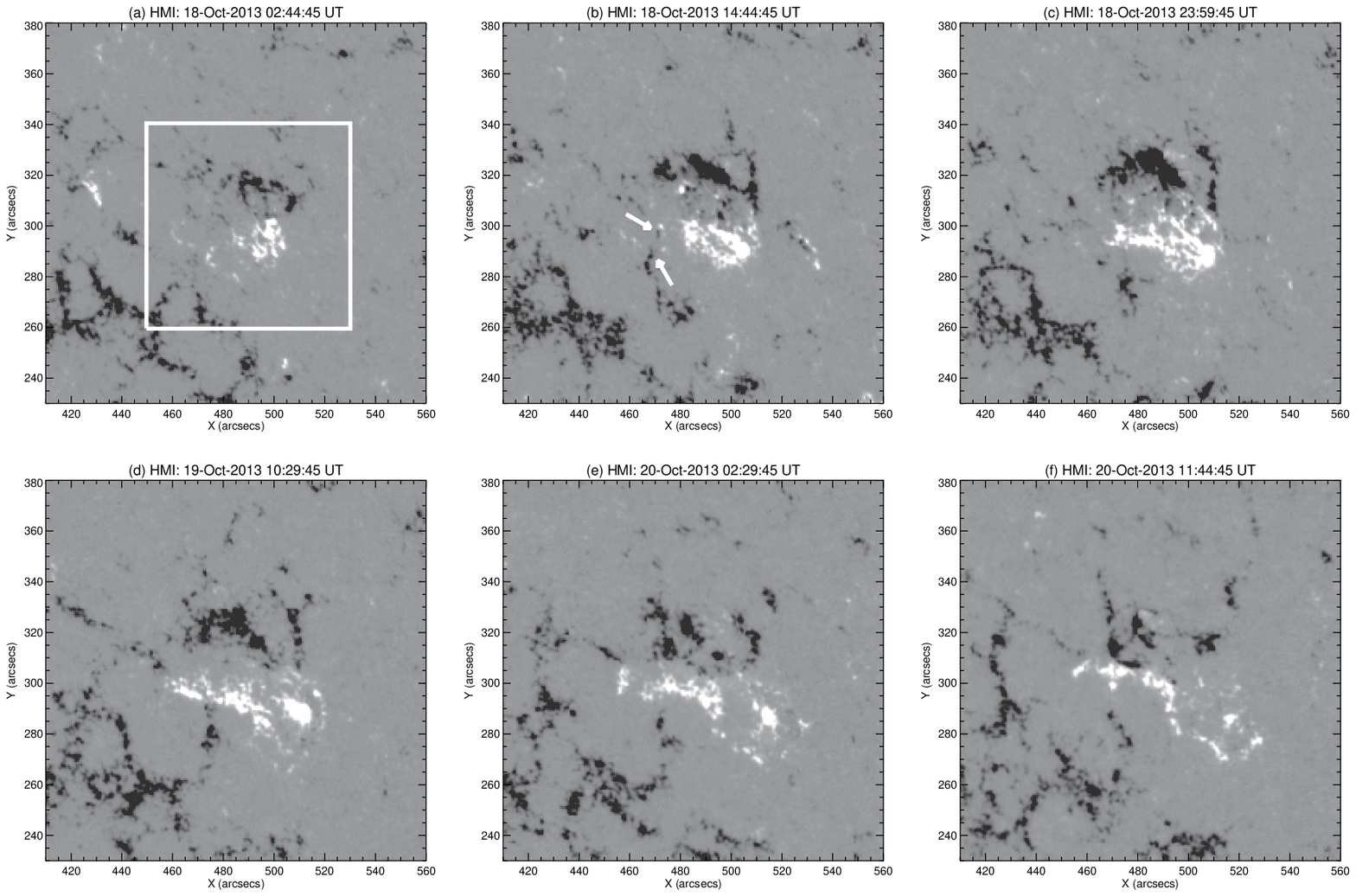}
\centerline{Figure~4}
\end{figure}
\clearpage

\clearpage
\pagestyle{empty}
\begin{figure}
\plotone{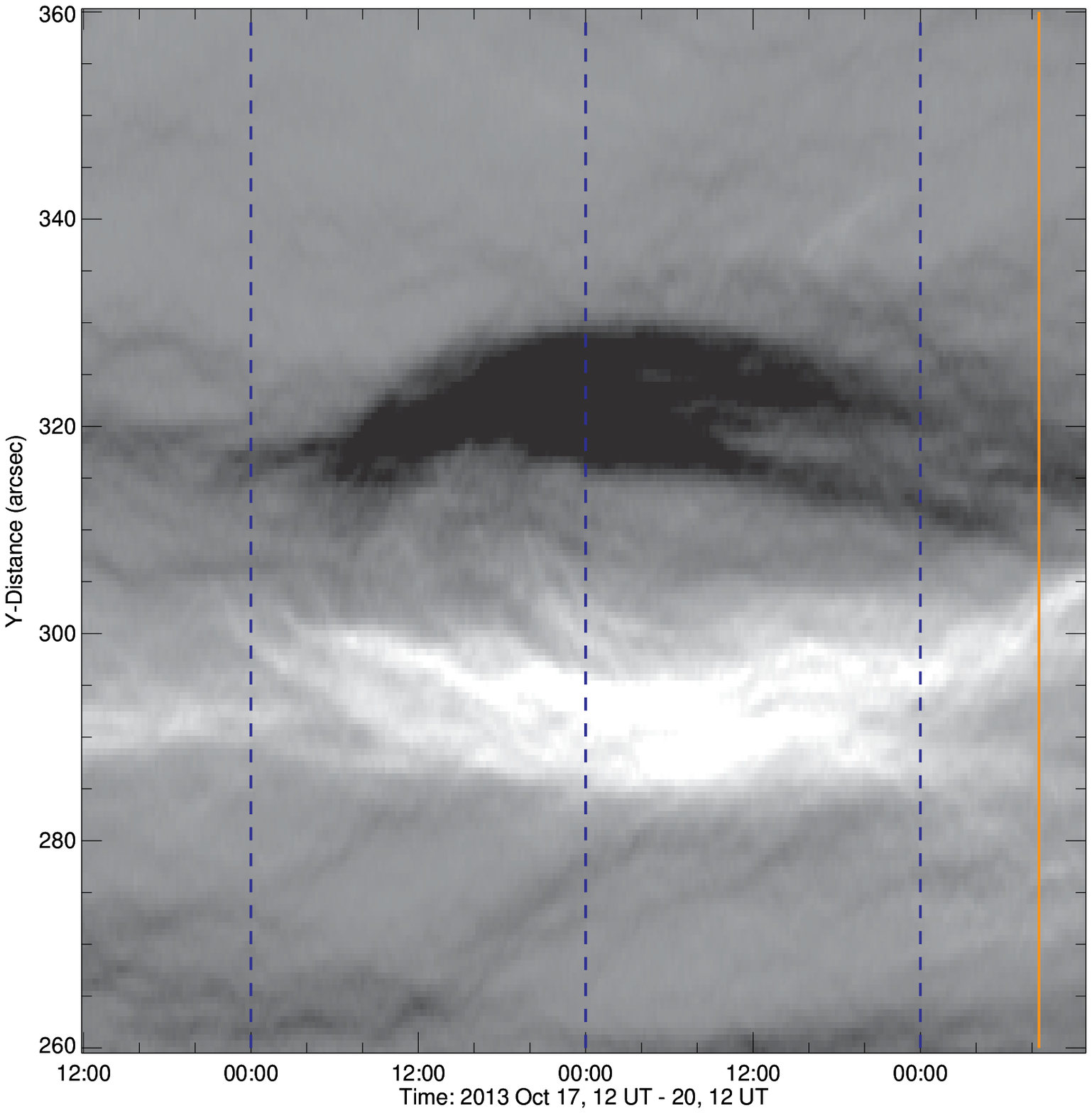}
\centerline{Figure~5}
\end{figure}
\clearpage

\clearpage
\pagestyle{empty}
\begin{figure}
\plotone{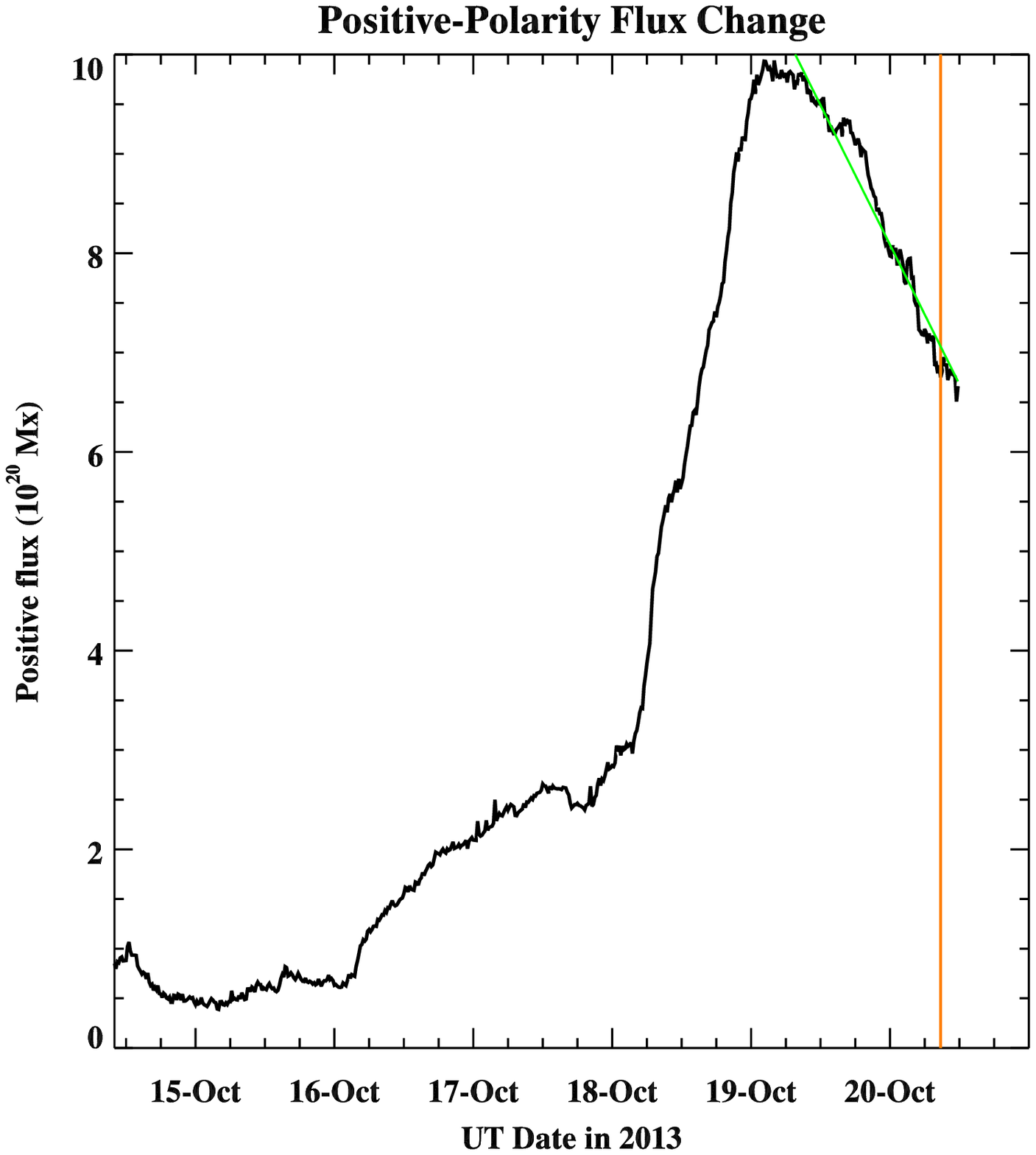}
\centerline{Figure~6}
\end{figure}
\clearpage

\clearpage
\pagestyle{empty}
\begin{figure}
\plotone{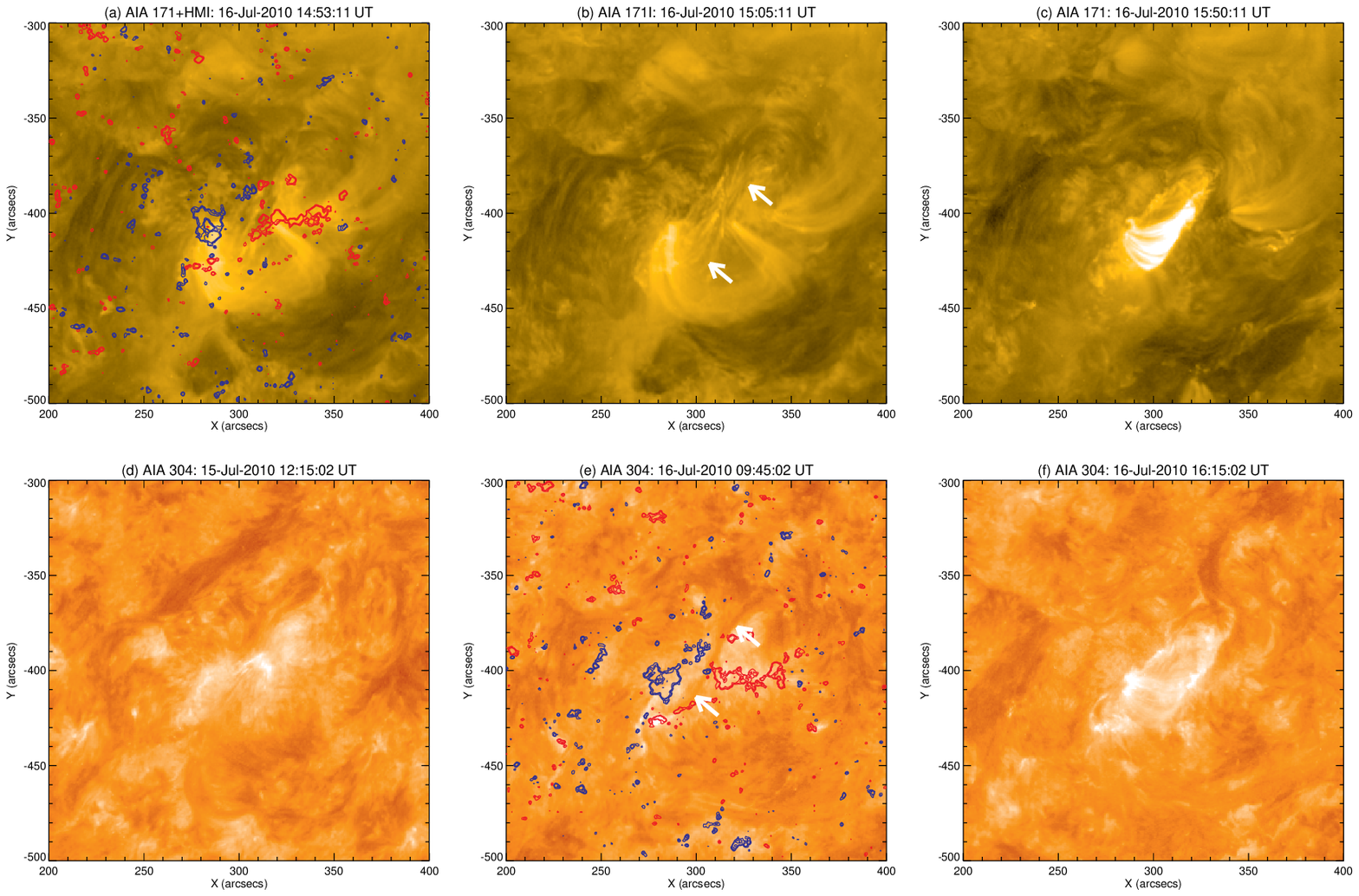}
\centerline{Figure~7}
\end{figure}
\clearpage

\clearpage
\pagestyle{empty}
\begin{figure}
\hspace*{0.5cm}\includegraphics[angle=0,scale=0.5]{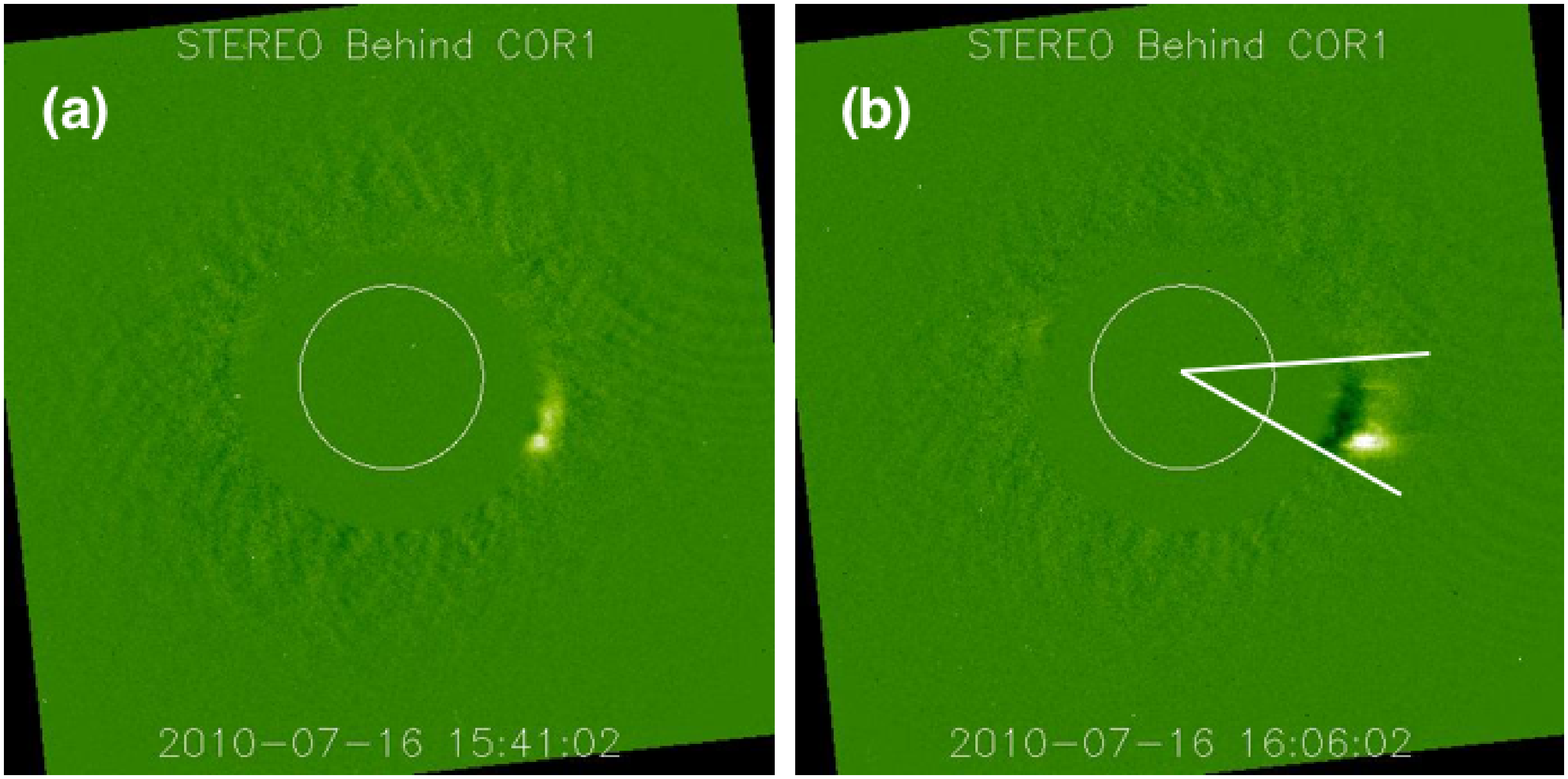}\vspace{1cm}
\centerline{Figure~8}
\end{figure}
\clearpage

\clearpage
\pagestyle{empty}
\begin{figure}
\plotone{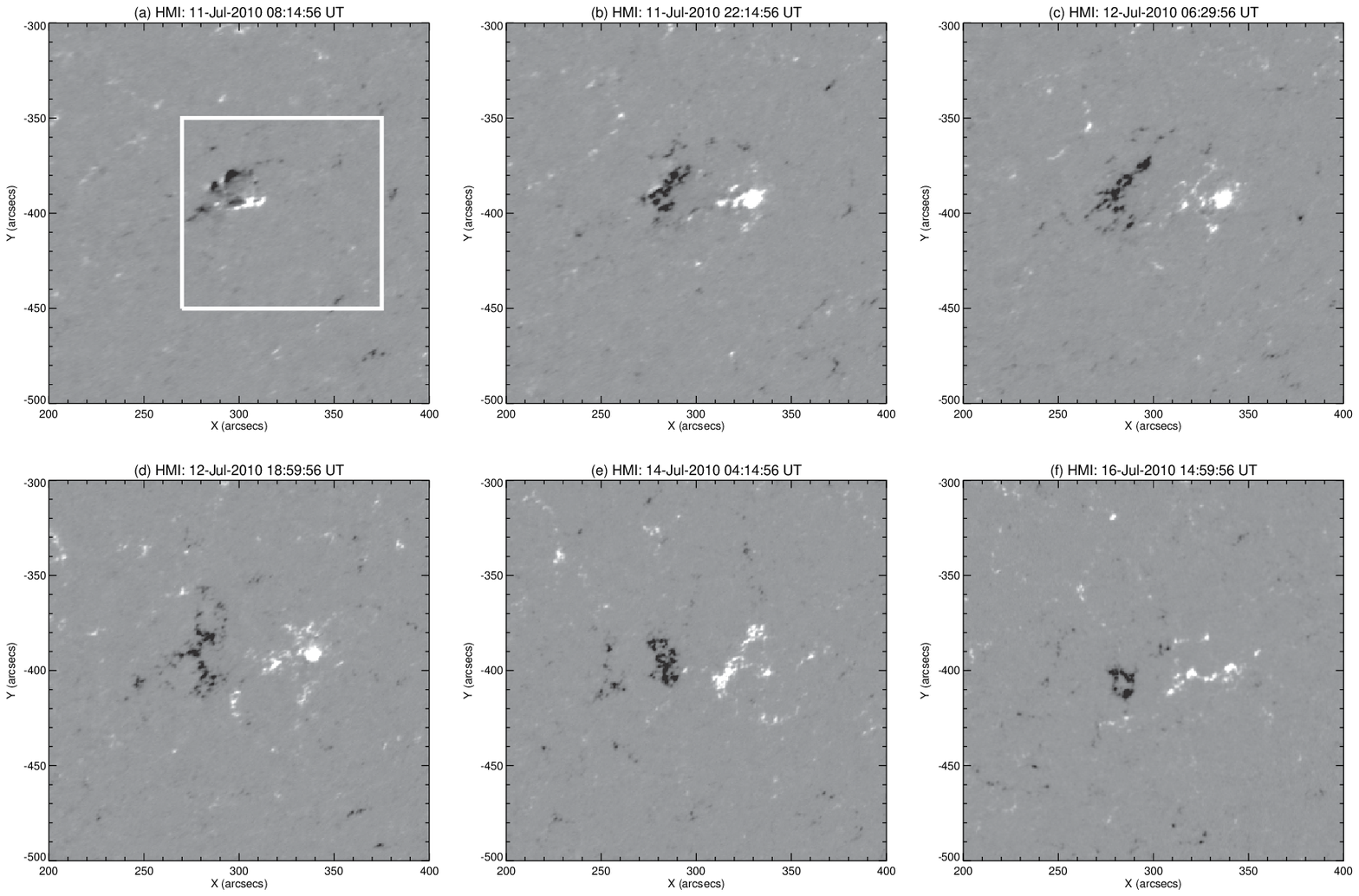}
\centerline{Figure~9}
\end{figure}
\clearpage

\clearpage
\pagestyle{empty}
\begin{figure}
\plotone{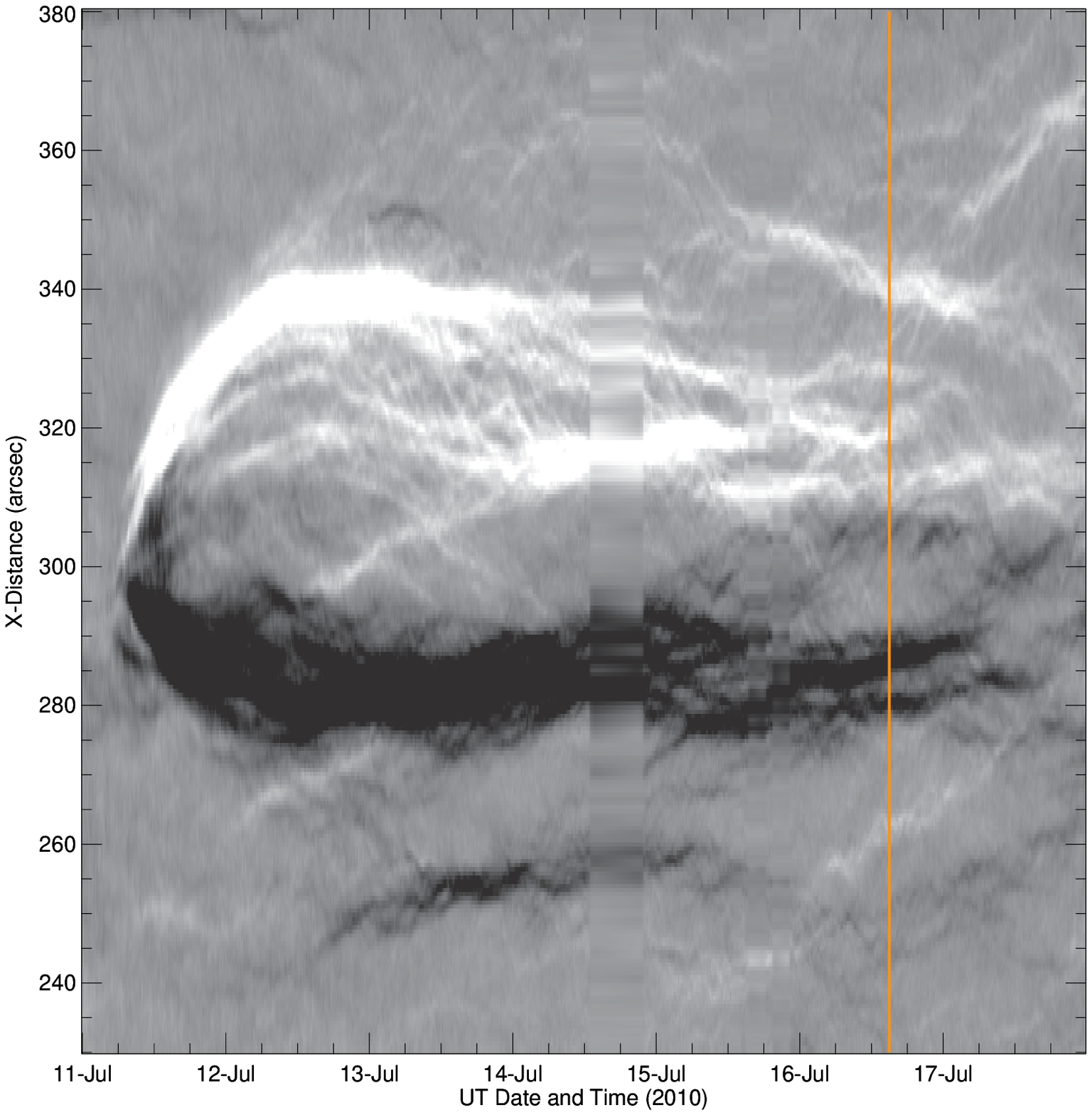}
\centerline{Figure~10}
\end{figure}
\clearpage

\clearpage
\pagestyle{empty}
\begin{figure}
\plotone{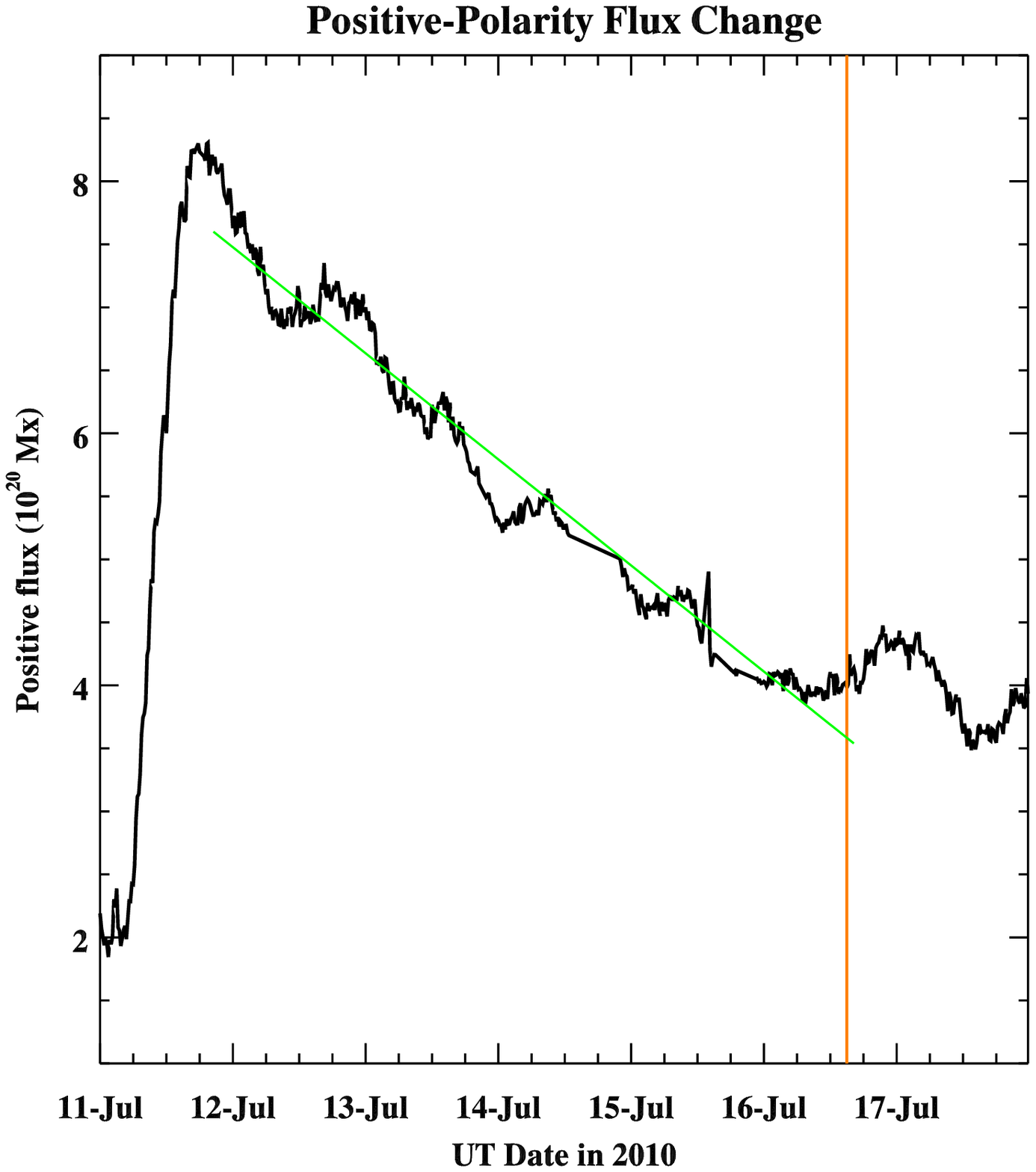}
\centerline{Figure~11}
\end{figure}
\clearpage

\end{document}